\documentclass[a4paper,nofootinbib,10pt]{revtex4}

\usepackage{verbatim}
\usepackage[dvips]{graphicx}
\usepackage{psfrag}
\usepackage[small,bf]{caption}
\usepackage{url}
\usepackage{amsfonts}
\usepackage{amsmath,amssymb}
\usepackage{slashed}
\usepackage{hyperref}

\newcommand{\lb}[0] { \left( }
\newcommand{\rb}[0] { \right) }
\newcommand{\g}[0] { \gamma }

\newcommand{\mn}[0] { {\mu \nu} }
\newcommand{\de}[0] { \delta }
\newcommand{\beqs} { \begin{eqnarray} }
\newcommand{\eeqs} { \end{eqnarray} }
\newcommand{\nn} {\nonumber}

\newcommand{\ep}[0] { \epsilon }
\newcommand{\noin}[0] {\noindent}
\newcommand{\sla}[1] {\slashed{#1}}
\newcommand{\im}[0] {\textrm{Im} \, }

\renewcommand{\eqref}[1]{(\ref{#1})}
\newcommand{\intsp}{\int \hspace{-0.1cm}}

\begin{document}

\title{Perturbative neutrino pair creation by an external source} 
\author{Hylke B.J. Koers}
\email{hkoers@nikhef.nl}
\affiliation{NIKHEF, P.O. Box 41882, 1009 DB Amsterdam, the Netherlands}
\affiliation{University of Amsterdam, Amsterdam, the Netherlands}

\begin{flushright}
  hep-ph/0409259  \\
  NIKHEF-2004-010 
\end{flushright}

\begin{abstract}
\noin We consider the rate of fermion-antifermion pair creation by an external field.
We derive a rate formula that is valid for a coupling with arbitrary vector and axial vector
components to first order in perturbation theory.
This is then applied to study the creation of neutrinos by nuclear matter, a problem with
astrophysical relevance. We present an estimate for the creation rate per unit volume, 
compare this to previous results and comment on the role
of the neutrino mass. \\

\noin PACS numbers: 13.15.+g, 95.30.Cq 
\end{abstract}

\maketitle

\section{Introduction}
\noin Starting with Schwinger's classical account \cite{Schwinger:1951nm} of electron-positron pair creation by an external
electric field, fermion pair creation has been the subject of continued interest.  A variety of pair
creation rates for specific external fields in quantum electrodynamics can be found in the literature, such as refs. \cite{Brezin:1970xf, Cornwall:1989qw, 
 Fried:2001ur, Grifols:1999ku, Hounkonnou:2000im, Kim:2000un, Kluger:1992gb, Lin:1998rn, Neville:1984va} and further references therein. 
The process exemplifies a true
quantum field theory phenomenon: the creation of particles from the vacuum.

Because neutrinos carry weak charge, one expects that an external
$Z$-boson field can produce neutrino-antineutrino pairs in a similar manner. The concept of an external $Z$-boson field
can be seen as arising from a distribution of nuclear matter (in the sense of ref. \cite{Kusenko:2001gb}).
Neutron stars are a prime example of such a matter distribution
and their neutrino emission by this mechanism
was studied  using non-perturbative
methods \cite{Kachelriess:1998cr,  Kusenko:2001gb, Loeb:1990nb}. Pair creation of neutrinos is also studied in relation to the stability of neutron stars, 
see Ref. \cite{Kiers:1997ty}  and references therein. Although Refs. \cite{Kachelriess:1998cr,  Kusenko:2001gb, Loeb:1990nb} find typical neutrino fluxes that are
too small to be
observable, we believe it is worthwhile to study such a relatively unexplored neutrino source from a different point
of view. In particular, we want to develop a method that is not limited to a specific source but allows us
to draw conclusions with a broad applicability. This can then be applied to study e.g. neutrino pair creation
by non-standard model weakly interacting particles or domain walls.\\

\noin In the present letter, we study the creation of neutrino pairs in a perturbative way.
 We present a first order computation of the pair creation rate per volume, with a
dynamical nuclear configuration acting as a source. The reasons for using perturbation theory are twofold. First, 
the axial coupling to the $Z$-boson prevents an easy generalization of
non-perturbative QED methods.
Second, non-perturbative methods generally consider a very specific source, or class of sources,
from the start. The perturbative method is more flexible in the sense that a specific source is folded in at the end. This allows
us to keep separate the physics of the pair creation process and that of a specific source.

In part, our computation was triggered by the results presented in ref. \cite{Kusenko:2001gb}, in which 
the creation of neutrinos by a 
time-dependent nuclear distribution is studied. One of the results in ref. \cite{Kusenko:2001gb} is
that the overall rate is proportional to the square of the neutrino mass. This implies that there can be no pair
creation of massless neutrinos. The question arises whether this is a manifestation of a general chiral 
suppression mechanism or a consequence of the specific source considered. 
We shall see that  the perturbative viewpoint contributes to a more complete understanding
of this effect. \\

\noin The paper is organized as follows. In section \ref{sect:pair} we discuss the theoretical background
of pair creation processes for fermions and introduce the relevant quantities. In section \ref{sec:twopoint}, we discuss
the perturbative computation.  The result is then applied to neutrinos in section
\ref{sect:nu} and we present our conclusions in section \ref{sect:con}.

\section{Pair creation physics}
\label{sect:pair}
\noin We study fermions that are coupled to an external source $j$. The interaction Lagrangian reads
\beqs
\label{eq:Lag}
\mathcal{L}_{\textrm{int}} = j_\mu(x) \,  \bar{\psi}(x) \Gamma^\mu \psi(x) \, .
\eeqs
The source is fully prescribed and has no further dynamics. We choose the coupling of the general form
\beqs
\Gamma^\mu = \gamma^\mu (c_V - c_A \g^5) \, ,
\eeqs
where $c_V$ ($c_A$) is the vector (axial vector) coefficient;
the coupling constant is absorbed in $j$. \\

\noin Following ref. \cite{Itzykson:1980rh}, we introduce
the overlap of asymptotic `in' and `out' vacua to describe the pair creation process:
\beqs
S_0(j) = \langle 0, \infty | 0, - \infty \rangle_j = \langle 0, \infty | S | 0,  \infty \rangle_j \, ,
\eeqs
where $S$ is the scattering operator and the subscript is a reminder that
a source is switched on and off adiabatically somewhere between $t=-\infty$
and $t=\infty$. The probability that a system that started in the vacuum state 
will remain in the vacuum state is then expressed \cite{Itzykson:1980rh}  as:
\beqs
\label{eq:defW}
| \langle 0, \infty | 0, - \infty \rangle_j |^2 = \exp{\lb -W \rb }
= \exp{ \lb - \intsp d^4 x \, w(x) \rb} \, .
\eeqs
For a positive $W$, this probability is  between zero and one which signals  a
non-zero probability for the creation of a fermion pair. 
Now suppose that $w(x)  = \bar{w} $ is constant. We can embed the system in a box of size
$V \times T$, write $W = \bar{w} V T$ and choose the box small enough
such that $W<1$:
\beqs
| \langle 0, \infty | 0, - \infty \rangle_j |^2 \simeq 1 -  \bar{w} V T \, ,
\eeqs
which supports the interpretation of the function $w(x)$ as
the probability per unit time and volume to create a pair at space-time
location $x$. Such a rate density is the physical quantity of interest.
For QED, the Schwinger formula  \cite{Schwinger:1951nm} states  that
for a photon field of the form $ A^\mu(x) = j^\mu(x) = (0,0,0,-e E t)$,
\beqs
\label{eq:Schw}
\bar{w} = \frac{\alpha E^2}{\pi^2} \sum_{n=1}^\infty \frac{1}{n^2} \exp{\lb - \frac{ n \pi m^2}{|e E|} \rb } \, ,
\eeqs
where $m$ is the electron mass. We mention that refs. \cite{Kachelriess:1998cr,  Loeb:1990nb}  conclude
that this result extends to the case of neutrino pair creation by a source of the same form. \\

\noin To compute the rate density, we use perturbative quantum field theory:
\beqs
 \langle 0, \infty | 0, - \infty \rangle_j = Z[j] = \exp{\lb i W[j] \rb} \, ,
\eeqs
where $W[j]$ is the generating functional of
connected $n$-point functions.\footnote{The use of $W$ and $W[j]$ may be confusing, but both symbols
are standard in the literature. The generating functional will always be denoted with its argument $j$.}
In this context, $W[j]$ is also the effective action  for the external field $j$ 
\cite{Itzykson:1980rh, Neville:1984va}.

The interaction Lagrangian \eqref{eq:Lag} only contains a  vertex that couples to the 
external field.
Therefore $W[j]$ represents an infinite sum of fermion loop diagrams, labeled
by the  number of vertices which are all connected to the external field.
In terms of $W$ that was introduced in eq. \eqref{eq:defW},
\beqs
W = 2 \textrm{ Im} \,  W[j] \, .
\eeqs
The fermion loop diagram with one external field vertex is zero
by momentum conservation, so the first non-zero contribution is from the
loop with two external field vertices, i.e. the two-point function. This is the object 
that we will compute in section \ref{sec:twopoint}. 
Its contribution  to  the pair creation rate $W_2$ is found
by folding in the sources according to the  formula\footnote{We use a metric tensor $g^\mn = \textrm{diag }(1,-1,-1,-1)$ throughout this paper.}
\beqs
\label{defP} W_2 = - \intsp \frac{d^4 p}{ (2 \pi)^4}j_\mu (p) j_\nu (-p)   \textrm{ Im}  \, \Sigma^\mn (p) \, , 
\eeqs
where $\Sigma^\mn$ represents the two-point function, with prefactors as 
chosen in eq. \eqref{Smn}. For time-like currents, $j_\mu(p) j_\nu(-p) \textrm{ Im} \,  \Sigma^\mn (p) < 0 $ since a probability cannot exceed one.
For a given $W$, the pair creation density follows by extracting the function $w(x)$. \\

\noin There has to be enough energy in the source to put two virtual 
particles on-shell. For the perturbative mechanism that we describe, this implies a threshold energy for 
the source insertions. This is in contrast to the non-perturbative effect, which can be thought of as an infinite sum of
loop diagrams with an increasing number of source insertions. 
This infinite amount of sources conspire to create a pair and the amount of energy per source insertion can be arbitrarily small.

For QED it is known that the real part of the sum of loop diagrams has a divergent structure, which can be used to extract non-perturbative results by performing a
Borel transformation \cite{Dunne:1999uy}. We do not know whether or not
a similar procedure can be applied in this more general situation.

\section{The two-point function}
\label{sec:twopoint}

\begin{figure}
\center
\includegraphics[width=5cm]{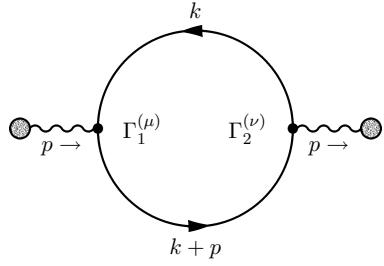}
\caption{Fermion loop diagram with two external sources attached. The external field
couples directly (i.e. without propagators) to the loop. \newline}
\label{figure:setuptwopoint}
\end{figure}

\noin The two-point function without external sources is
transcribed from figure \ref{figure:setuptwopoint}.  We find that, in dimensional
regularization with $n = 4 - \ep$,
\beqs
\label{Smn} \Sigma^\mn (p) = - i \mu^{(4-n)} \intsp \frac{d^n k}{(2 \pi)^n}
\frac{ \textrm{tr}\, { \left[ \lb \sla{k}+m \rb \Gamma^\mu  \lb \sla{k}+\sla{p}+ m \rb \Gamma^\nu  \right]}}
{( k^2 - m^2 + i \ep ) ( ( k + p )^2 - m^2 + i \ep )} \, ,
\eeqs
where $m$ is the fermion mass.
From eq. \eqref{defP}, we are interested in the imaginary part of this expression, which is finite. 
Note that we integrate over the fermion momentum; in the source's rest frame (where the
particles are created back to back), the fermion and the antifermion each carry half of the energy.

Expression \eqref{Smn} is reduced to a linear combination of scalar
integrals in the fashion of Passarino-Veltman \cite{Passarino:1979jh}. A
series expansion in
$\ep$ reveals the divergent structure, and after  some 
algebra 
 the problem depends only
 on the one- and two-point scalar integrals. The one-point scalar integral is real,
 the two-point integral develops an imaginary part if $ p^2 > 4m^2$ which means there should be enough energy in the source to
create two fermions. If this is not satisfied,  $\Sigma^\mn$ is purely real
and there is no pair creation. 
The final result is the following expression:
\begin{subequations}
\label{Smngen}
\beqs
\textrm{Im} \, \Sigma^\mn (p) & = & \frac{1}{16 \pi^2} \left[ \lb c_V^2 - c_A^2 \rb \Sigma^\mn_I (p) + \lb c_V^2 + c_A^2  \rb 
\Sigma^\mn_{II} (p) \right]  \theta(p^2 - 4 m^2) \, ,  \\
\Sigma^\mn_I (p) & = & 4 \pi  m^2  \sqrt{1 -\frac{4 m^2}{p^2}} g^\mn \, , \\
\Sigma^\mn_{II} (p) & = & \frac{4}{3} \pi   \lb  p^2 g^\mn  - p^\mu p^\nu  - \frac{2 m^2}{p^2} p^\mu p^\nu - m^2 g^\mn \rb \sqrt{1 -\frac{4 m^2}{p^2}} \, .
\eeqs
\end{subequations}
For some typical values of the parameters $c_V$ and $c_A$, this result can be 
compared to the literature  \cite{Chang:1982qq,Itzykson:1980rh}. \\

\noin From expressions \eqref{Smngen} we observe that  for massless fermions \mbox{$\Sigma_I^\mn = 0$}, so that only
the second term contributes. This means that the physics is qualitatively
insensitive to different choices of $c_V$ and $c_A$; only the square sum
is quantitatively important. We conclude that the difference between the two-point functions with two
different normalized sets of couplings (e.g. purely vector, purely axial vector)
is proportional to $m^2$. 

The contribution due to the three-point diagram should be interpreted
with care. In QED it vanishes by Furry's theorem, but for axial
couplings it contributes to the axial anomaly. This means one should
verify that the final result does not depend on the regularization
procedure. For the present calculation, this is not an issue.

\section{Neutrino pair creation to first order}
\label{sect:nu}

\subsection{The general case}
\noin We specialize to neutrino pair creation by putting $c_V = c_A =1/2 $ in the expression for
the two-point function \eqref{Smngen}. Combining  eqs. \eqref{defP} and \eqref{Smngen}, we find
\begin{subequations}
\label{probnu}
\beqs
 W_2   & = & - \frac{1}{24 \pi} \intsp \frac{d^4 p}{(2 \pi)^4} \,  \theta (p^2 -4 m^2)\sqrt{1 -  \frac{4m^2}{p^2}} 
\left[ \mathcal{F}_0 (p,j) +  m^2 \mathcal{F}_1 (p,j)  \right]  \, , \\
\mathcal{F}_0 (p,j) &  = & p^2 \left[ j(p) \cdot j(-p) \right]
-  \left[ p \cdot j(p) \rb \lb p \cdot j(-p) \right]  \, , \\
 m^2 \mathcal{F}_1 (p,j) &  = & - m^2 \left[ j(p) \cdot j(-p) \right]
- \frac{2 m^2}{p^2}  \left[ p \cdot j(p) \rb \lb p \cdot j(-p) \right]  \, .
\eeqs
\end{subequations}
Without loss of generality, we  consider a source with a density component and a spatial current in the $ \hat{z} $ direction:
\beqs
\label{eq:sourcegen}
j_\mu (p) = (j_0 (p), 0,0, j_3(p) ) \, , \qquad p_\mu=(E,\vec{p}_T,p_3) \, .
\eeqs
Here $p_\mu$ labels the energy and momentum of the source. Though the 
current is directed in the $\hat{z}$ direction, we allow for a
dependence on the transverse direction by leaving $\vec{p}_T$ unspecified.
The two terms in \eqref{probnu} can be written as
\begin{subequations}
\label{probnu2}
\beqs
  \mathcal{F}_0 (p,j) & = &  - {\vec{p}_T}^2 \lb  | j_0 |^2 - |j_3|^2 \rb - | E j_3  - p_3 j_0 |^2 \, , \\
\nn m^2 \mathcal{F}_1 (p,j) & = &
 - \frac{2 m^2}{E^2 - {\vec{p}_T}^2- {p_3}^2} \lb E^2  |j_0|^2 + {p_3}^2 |j_3|^2 -E p_3 \lb j_0 j_3^* + j_0^*
j_3 \rb \rb  \\
 & & \qquad  -m^2 \lb | j_0 |^2 - |j_3|^2 \rb \, . 
\eeqs
\end{subequations}
We do not simplify these equations further, as we do not want to constrain the properties of the source.

It is instructive to analyze the massless limit in more detail.  In this case only $\mathcal{F}_0$ in \eqref{probnu} contributes, so that
\beqs
 W_2 \, (m =0)  & = & - \frac{1}{24 \pi} \intsp \frac{d^4 p}{(2 \pi)^4} 
\left[ p^2 \lb j(p) \cdot j(-p) \rb
-  \lb p \cdot j(p) \rb \lb p \cdot j(-p) \rb \right] \, .
\eeqs
In analogy with QED, we introduce a field strength $F_\mn(p) = i p_\nu j_\mu (p) - i p_\mu j_\nu (p)$ and its `electric' and `magnetic'
components $E_i$ and $B_i$ and find:
\begin{subequations}
\label{W2res}
\beqs
 W_2 \, (m=0) & =& - \frac{1}{48 \pi} \intsp \frac{d^4 p}{(2 \pi)^4} 
\left[ F_\mn (p) F^\mn (-p) \right] \\
& = &   \frac{1}{24 \pi} \intsp \frac{d^4 p}{(2 \pi)^4}   \left[  E_i(p) E_i(-p) - B_i(p) B_i (-p)  \right] \, .
\eeqs
\end{subequations}
This is exactly half of the QED result \cite{Itzykson:1980rh} if we insert a factor $e^2$ from the coupling constants, which reflects the discussion in the previous section. In electrodynamics,
$\vec{E}$ and $\vec{B}$ are the physical electric and magnetic fields and one can  go to a frame in which $\vec{B}=0$.
Then eq. \eqref{W2res} yields a positive result from which we conclude that the creation of massless particles by the two-point mechanism
is in general possible.
Eq. \eqref{W2res} is consistent with the massless limit of the first-order effective action in an axial background
that was computed in ref. \cite{Maroto:1998zc}.

It is interesting to compare this result to the creation of neutrinos by an
external electromagnetic field  as computed in ref. \cite{Gies:2000wc}. In that case, the pair creation rate
is proportional to $m^2$ and depends on the electromagnetic invariant $\vec{E} \cdot \vec{B}$.

\subsection{The time-dependent density}
\noin We
consider a time-dependent distribution of nuclear matter, described by the following source term:\footnote{This
source originates from an effective four-fermion description, see ref. \cite{Kusenko:2001gb}.
Note that  $j^\mu$ contains the axial current; since the neutrons
are massive, axial symmetry is broken and the current need not be divergence-free.}
\begin{subequations}
\label{tdd}
\beqs
 j_\mu (t) &=& G_F / \sqrt{2} \,  \langle \overline{n} \g_\mu (1-\g^5) n \rangle = \left( j_0 (t),0,0,0 \right) \, , \\ 
\label{eq:tdd} j_0(t)& =& \frac{G_F}{\sqrt{2}} \, n_N(t) \, ,
\eeqs
\end{subequations}
where $n_N$ is the number density of the nuclear matter distribution and $G_F$ is Fermi's constant.
This is the specific background that we refer to as a
time-dependent density.
Our main motivation for this source is to compare the perturbative results with the non-perturbative results of ref. \cite{Kusenko:2001gb}.

For simplicity (and because any source can be decomposed into a trigonometric sum) we assume a monochromatic source: \mbox{$ j_0 (t) = E_0 \cos{\omega t}$}. In Fourier space,  this is
\begin{subequations}
\label{tdd2}
\beqs
j_0 (p) &= & \frac{E_0}{2}  (2 \pi)^4    \de  (\vec{p}) \left[ \de(E-\omega) + \de(E+\omega)   \right] \, , \\
E_0 &= & \frac{G_F}{\sqrt{2}} n_N (0) \, .
\eeqs 
\end{subequations}
Inserting the source \eqref{tdd2} into eq. \eqref{defP} results in products  of delta functions. We employ a box normalization procedure
to reduce these to a single delta function and a factor $V \times T$ and find
\beqs
\label{box1} W_2  = - V T \frac{{G_F}^2  \lb n_N \rb^2 }{8} \left[  \im  \Sigma^{0 0} (\omega; \vec{p}=0) + \im   \Sigma^{0 0} (-\omega; \vec{p}=0) \right] \, .
\eeqs
Using eq. \eqref{Smngen}, with $c_V = c_A = 1/2$, we see
\beqs
 \textrm{Im} \, \Sigma^{0 0}  (\pm \omega; \vec{p}=0) = - \frac{ m^2}{8 \pi}  \sqrt{1 - \frac{4 m^2 }{\omega^2} } \, ,
\eeqs
leading to the following pair creation probability per unit time and volume:
\beqs
\label{tddrate} \bar{w}_2 =  \frac{W_2}{V T} 
= \frac{m^2}{32 \pi}  \sqrt{1 - \frac{4 m^2 }{\omega^2} } {G_F}^2  {n_N}^2 \, . 
\eeqs
The rate density scales with the square of the nuclear density, 
as expected for the two-point mechanism. 

We see that the rate is proportional to $m^2$, which could have been anticipated
from eq. \eqref{probnu2} because the time-dependent density  \eqref{tdd}
is characterized by $j_3 = \vec{p}_T = p_3 =0$ so that $\mathcal{F}_0 (p,j)= 0$.
Ref. \cite{Kusenko:2001gb} also finds the $m^2$ proportionality for sources with a time-dependent current
in the $\hat{z}$-direction. Equation \eqref{probnu2} suggests that such sources can contribute
to first order for a zero neutrino mass. \\

\noin  To derive an order-of-magnitude  estimate for the 
number of created neutrinos per unit volume per unit time, we take the square root factor in eq.
\eqref{tddrate} of order unity, use a neutrino mass of $0.1$ eV and assume a `reduced density' $G_F n_N / \sqrt{2} \sim  1$ eV, such as in a neutron star \cite{Kusenko:2001gb}:
\beqs
\bar{w}_2 =  \frac{ \lb 0.1 \, \textrm{eV} \rb^2}{32 \pi} \lb 2 \, \textrm{eV}^2  \rb \sim  10^{-4} \textrm{eV}^4 \sim 10^{26} \textrm{ s}^{-1} \textrm{ cm}^{-3} \, .
\eeqs
At the pair creation threshold, this corresponds to an energy output of
order $ 10^{13}  \textrm{ erg} \textrm{ cm}^{-3} \textrm{ s}^{-1} $. Ref. 
\cite{Kusenko:2001gb} estimates the energy output of neutrinos that are created non-perturbatively
by an oscillating neutron star to be of order $ 10^{3}  \textrm{ erg} \textrm{ cm}^{-3} \textrm{ s}^{-1} $. 
However, these numbers should not be compared because the (realistic) driving frequency that is considered in \cite{Kusenko:2001gb} is so low that the perturbative
mechanism is not operational. 

As follows from eq. \eqref{Smngen}, there can only be pair creation by the two-point mechanism if
$\omega^2 > 4 m^2$. With a neutrino mass of $0.1$ eV, the creation of a neutrino-antineutrino pair requires a driving frequency of at least
$ 3 \cdot 10^{14} \, \textrm{Hz}$. The coherence length of such a system is roughly $10^{-4}$ cm, so
it is not very feasible to look for an oscillating astrophysical object
that would produce an appreciable amount of neutrinos with this mechanism.
However, the value of our computation lies in its general applicability. We are
not limited to this particular type of sources, and we believe it may be interesting to study sources of a more
transient nature such as a forming neutron star. Alternatively, one could consider weakly interacting 
particles beyond the standard model or domain walls  as a source.

\section{Conclusion}
\label{sect:con}
\noin We have described pair creation of fermions by an external field
to first order in perturbation theory and found the contribution by the two-point mechanism for a general coupling. Our main result is eq. \eqref{Smngen}, which should be
interpreted in the context of eq. \eqref{defP}. We observe that at this order
in perturbation theory, the difference in pair creation rates between two sets of normalized coupling coefficients
\mbox{\{$c_V$, $c_A$\}} is proportional to the square of the fermion mass.

For the case of neutrino pair creation by a distribution of nuclear matter, we have derived
expressions \eqref{probnu} and \eqref{probnu2}. From this result
we observe that, to first order, neutrino pair creation is possible with a suitable source if
neutrinos would have been massless particles. We then considered pair creation of neutrinos 
by the time-dependent density of eq. \eqref{tdd}.
For this specific source, we conclude that the production rate due to the two-point 
contribution \eqref{tddrate}
is proportional to the square of the neutrino mass. This conclusion is in qualitative
agreement with the non-perturbative result derived in \cite{Kusenko:2001gb}.

The method that we presented in this paper is suitable to study different types of sources and we intend to do so
in the future. \\

\begin{acknowledgments}
\noin The author wants to thank Karel Gaemers, Jan-Willem van Holten, Eric Laenen and  Marieke Postma for very useful discussions
and suggestions.
\end{acknowledgments}

 \bibliography{npp}

\end{document}